\documentstyle[fleqn,run2col,epsfig]{article}

\begin{document}
%
\thispagestyle{empty}
\onecolumn
\begin{flushright}
\large
YITP-00-04 \\
February 2000
\end{flushright}
\vspace{1.2cm}

\renewcommand{\thefootnote}{\fnsymbol{footnote}}
\setcounter{footnote}{1}
\begin{center}  
\begin{LARGE} 
\begin{bf}
Soft-Gluon Resummation and PDF Theory 
Uncertainties\footnote{
\begin{large} 
Contribution to the proceedings of the workshop
``Physics at Run II'', Fermilab, 1999.
\end{large}}

\end{bf}  
\end{LARGE}

\vspace*{1.8cm}
{\Large George~Sterman and  Werner~Vogelsang}

\vspace*{4mm}

\begin{large}
{C.N.\ Yang Institute for Theoretical Physics, State University of
New York, Stony Brook}

\vspace*{2mm}
Stony Brook, NY 11794-3840, U.S.A.\\[3pt]
\vspace*{2mm}
E-mails: {\tt sterman@insti.physics.sunysb.edu,
vogelsan@insti.physics.sunysb.edu}
\end{large}
\vspace*{2.cm}

{\Large \bf Abstract}
\end{center}

\noindent
Parton distribution functions are determined
by the comparison of finite-order calculations with data.
We briefly discuss the interplay of higher order corrections
and PDF determinations, and the use of soft-gluon resummation in
global fits.

\setcounter{page}{0}
\renewcommand{\thefootnote}{\arabic{footnote}}
\setcounter{footnote}{0}
\normalsize
\newpage

\def\beq{\begin{equation}}
\def\eeq{\end{equation}}
\def\beqa{\begin{eqnarray}}
\def\eeqa{\end{eqnarray}}
\def\a {{\rm f}}
\def\e{r}
\def\g{\xi}
\def\w{\rho}
\def\y{\eta}
\def\Qg{Q_{\rm gap}}

\newcommand{\ttbs}{\char'134}
\newcommand{\AmS}{{\protect\the\textfont2
   A\kern-.1667em\lower.5ex\hbox{M}\kern-.125emS}}

\hyphenation{author another created financial paper re-commend-ed}

\title{Soft-Gluon Resummation and PDF Theory Uncertainties}

\author{George Sterman and Werner Vogelsang\address{C.N.\ Yang 
Institute for Theoretical Physics \\
SUNY at Stony Brook,
Stony Brook, NY 11794-3840, USA}%
         \thanks{This work was supported in part by the National 
Science Foundation, grant PHY9722101.}}

\begin{abstract}
Parton distribution functions are determined
by the comparison of finite-order calculations with data.
We briefly discuss the interplay of higher order corrections
and PDF determinations, and the use of soft-gluon resummation in
global fits.
\end{abstract}

\maketitle

\section{FACTORIZATION \& THE NLO MODEL}

A generic inclusive
cross section for the process $A+B\rightarrow F+X$
with observed final-state system $F$, of total mass $Q$, can be expressed as
\beqa
Q^4\, {d\sigma_{AB\rightarrow FX}\over dQ^2}
&=&
\phi_{a/A}(x_a,\mu^2)\; \otimes\;  \phi_{b/B}(x_b,\mu^2)
\nonumber\\
&\ &  \quad\quad  \otimes\;
{\hat \sigma}_{ab\rightarrow FX}\left (z,Q,\mu\right)\, ,
\label{basicfact}
\eeqa
with $z=Q^2/x_ax_bS$. The
$\hat{\sigma}_{ab}$ are partonic hard-scattering functions,
$
{\hat\sigma}=\sigma_{\rm Born}+(\alpha_s(\mu^2)/\pi){\hat\sigma}^{(1)}
+\dots\, .
$
They are known to NLO for most processes in the
standard model and its popular extensions.  Corrections
begin with higher, uncalculated orders in the hard scattering,
which respect  the
form of Eq.\ (\ref{basicfact}).
The discussion is simplified in terms of moments
with respect to $\tau=Q^2/S$,
\beqa
\tilde\sigma_{AB\rightarrow FX} &=& \int_0^1 d\tau\;
\tau^{N-1}\ Q^4\, {d\sigma_{AB\rightarrow FX} / dQ^2}
\nonumber\\
&& \hspace{-23mm}
= \sum_{a,b}\, \tilde \phi_{a/A}(N,\mu^2)\; \tilde\sigma_{ab\rightarrow 
FX}(N,Q,\mu)\;
\tilde\phi_{b/B}(N,\mu^2)\, ,
\label{momentact}
\eeqa
where the moments of the $\phi$'s and 
$\hat\sigma_{ab\rightarrow FX}$ are defined similarly.

Eqs.\ (\ref{basicfact}) and (\ref{momentact}) are starting-points for
both the determination and the application of
parton distribution functions (PDFs), $\phi_{i/H}$,
using 1-loop $\hat\sigma$'s \cite{mrst99,grv98,cteq5}
We may think of this collective enterprise as an ``NLO model" for
the PDFs, and for hadronic hard scattering in general.
For precision applications we ask how well we
really know the PDFs \cite{disPDF,uncerglue,bayseanpdfs}.
Partly this is a question of how
well data constrain them, and partly it is a question of
how well we {\it could} know them, given finite-order
calculations in Eqs.\ (\ref{basicfact}) and (\ref{momentact}).
We will not attempt here to assign error estimates
to theory.  We hope, however, to give a sense of how
to distinguish ambiguity from uncertainty, and how
our partial knowledge of higher orders can reduce
the latter.

\section{UNCERTAINTIES, SCHEMES \& SCALES}

It is not obvious how to quantify
a ``theoretical uncertainty",
since the idea seems to require us to estimate corrections that
we haven't yet calculated.  We do not think an
unequivocal definition is possible, but we can try at least to
clarify the concept, by
considering a hypothetical set of nucleon PDFs
determined from DIS data alone \cite{disPDF}.  To make such a
determination, we would invoke isospin symmetry
to reduce the set of PDF's to those of the proton, $\phi_{a/P}$,
and then measure a set of singlet and
nonsinglet structure functions, which we denote $F^{(i)}$.
Each factorized structure function may be written
in moment space as
\beq
\tilde F^{(i)}(N,Q) = \sum_a \tilde C^{(i)}_a(N,Q,\mu)\, \tilde 
\phi_{a/P}(N,\mu^2)\, ,
\label{disfact}
\eeq
in terms of which we may solve for the parton
distributions by inverting the matrix $\tilde{C}$,
\beq
\tilde \phi_{a/P}(N,\mu^2) = \sum_i \, \tilde C^{-1}{}^{(i)}_a(N,Q,\mu)\
\tilde F^{(i)}(N,Q)\, .
\label{solvephi}
\eeq
With ``perfect" $\tilde F$'s at fixed $Q$,
and with a specific approximation for the coefficient functions,
we could solve for the moment-space distributions numerically,
without the need of a parameterization.  In a world of
perfect data, but of incompletely known coefficient functions, uncertainties
in the parton distributions would be entirely due to
the ``theoretical" uncertainties of the $C$'s:
\beq
\delta \tilde \phi_{a/P}(N,\mu) = \sum_i \, \delta \tilde 
C^{-1}{}^{(i)}_a(N,Q,\mu)\
\tilde F^{(i)}(N,Q)\, .
\label{delphidelC}
\eeq
Our question now becomes, how well do we know the $C$'s?
In fact this is a subtle question, because
the coefficient functions depend on choices of scheme
and scale.  

Factorization schemes are procedures
for defining coefficient functions perturbatively.
For example, choosing for $F_2$ the LO (quark) coefficient function
in Eq.\ (\ref{solvephi}) defines a DIS scheme
(with $\tilde{C}$ independent of $\mu$, which is then to
be taken as $Q$ in $\tilde{\phi}$).  Computing
the $C$'s from partonic cross sections by minimal subtraction
to NLO defines an NLO $\overline{\rm MS}$ scheme, and so on.
Once the choices of $C$'s and $\mu$
are made, the PDF's are defined
uniquely.  

Evolution in an $\overline{\rm MS}$
or related scheme, enters through
\beqa
\mu{d\over d\mu} \tilde \phi_{a/H}(N,\mu^2) \hspace*{-2mm} 
&=&\hspace*{-2mm}  -  \Gamma_{ab}(N,
\alpha_s(\mu^2))\, 
\tilde \phi_{b/H}(N,\mu^2) \nonumber \\
\mu{d\over d\mu}\tilde C^{(i)}_c(N,Q,\mu)\hspace*{-2mm}  &=&
\hspace*{-2mm}   \tilde C^{(i)}_d(N,Q,\mu)\, 
\Gamma_{dc}(N,\alpha_s(\mu^2))\, . 
\label{evol}
\eeqa
In principle, by Eq.\ (\ref{evol}), the scale-dependence of the $C_a^{(i)}$
exactly cancels that of the PDFs in Eq.\ (\ref{disfact}) and, by
extension, in Eq.\ (\ref{basicfact}).
This cancelation, however, requires that each $C$
and the anomalous dimensions $\Gamma$ be known
to all orders in perturbation theory.

To eliminate $\mu$-dependence
up to order $\alpha_s^{n+1}$, we need $\hat\sigma$ to order
$\alpha_s^n$ and the $\Gamma_{ab}$ to $\alpha_s^{n+1}$.
One-loop (NLO) QCD corrections to hard scattering require
two-loop splitting functions, which are known.  The complete
form of the  NNLO splitting functions, is still somewhere over
the horizon \cite{nnlo}.  Even when these are
known, it will take some time before more than a few hadronic
hard scattering functions are known at NNLO.

We can clarify the role of higher orders by relating  structure functions
at two scales,$Q_0$ and $Q$. Once we have measured
$F(N,Q_0)$, we may predict $F(N,Q)$ in terms of the relevant
anomalous dimensions and coefficient functions by
\beqa
F(N,Q) &=& F(N,Q_0)\ {\rm e}^{\int_{Q_0}^{Q} {d\mu'\over\mu'}\,
\Gamma(N,\alpha_s(\mu'{}^2))}
\nonumber\\
&\ & \quad \times \left [ {\tilde{C}(N,Q,Q)\over 
\tilde{C}(N,Q_0,Q_0)}\right]\, .
\label{predict}
\eeqa
This prediction, formally independent of PDFs
{\it and} independent of the factorization scale,
has corrections from the next, still uncalculated
order in the anomalous dimension and in the ratio of
coefficient functions.  The asymptotic freedom of QCD
gives a special role to LO: only the one-loop contribution
to $\Gamma$ diverges with $Q$ in the exponent, and contributes
to the leading, logarithmic scale breaking.  NLO
corrections already decrease as the inverse of the logarithm
of $Q$, NNLO as two powers of the log.  Thus, the theory
is self-regulating towards high energy,
where dependence on uncalculated pieces in
the coefficients and anomalous dimensions becomes less and
less important.

The general successes of the NLO model strongly
suggest that relations like (\ref{predict})
are well-satisfied for a wide range of observables
and values of $N$ (or $x$) in DIS and other processes.
This does not mean, however, that we have no knowledge of,
or use for, information from higher orders. In particular, 
near $x=1$ PDFs are rather poorly known \cite{cteqlargex}.
At the same time, the 
ratio of $C$'s depends on $N$, and if $\alpha_s\ln N$ is large, it becomes
important to control higher-order dependence on $\ln N$. This is a task 
usually referred to as resummation, to which we now turn.

\section{RESUMMATION}

\begin{figure}[t]
\epsfig{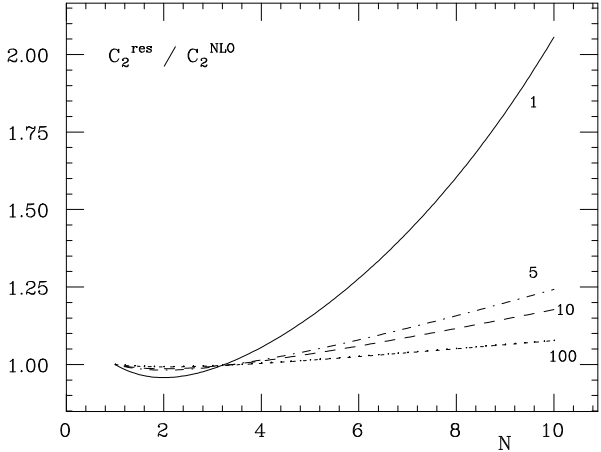}
\vspace{-0.6cm}
\caption{Ratio of Mellin-$N$ moments of resummed and NLO 
$\overline{\rm MS}$-scheme quark coefficient functions for $F_2$.
The numbers denote the value of $Q^2$ in GeV$^2$. We have
chosen $\mu=Q$.}
\label{cratio}
\end{figure}

Let us continue our discussion of DIS, describing
what is known about the $N$-dependence of the coefficient
functions $C$, as a step toward understanding
the role of higher orders.  Specializing again for simplicity to
nonsinglet or valence, the resummed coefficient function may
be written as \cite{oldDY,Cea1PI}
\beq
\tilde C^{\rm res}(N,Q,\mu) = \tilde C^{\rm NLO}_{sub}(N,Q,\mu)+
C_{\delta}^{\rm DIS} \, {\rm e}^{E_{\rm DIS}(N,Q,\mu)},
\label{Cresum}
\eeq
where ``{\it sub}" implies a subtraction on
$\tilde{C}^{\rm NLO}$ to keep $\tilde{C}^{\rm res}$ exact at order
$\alpha_s$, and where $C_{\delta}^{\rm DIS}$ corresponds to the NLO 
$N$-independent (``hard virtual'') terms.  
The exponent resums logarithms of $N$:
\beqa
E_{\rm DIS}(N,Q,\mu) &=& \\
&\ & \hspace{-30mm}
\int_{Q^2/\bar{N}}^{\mu^2}{d\mu'{}^2\over\mu'{}^2}
\Big[ A(\alpha_s(\mu'{}^2))\ln( \bar{N}\mu'{}^2/Q^2)
  + B(\alpha_s(\mu'{}^2))\Big], \nonumber
\label{Edisdef}
\eeqa
with $\bar{N} \equiv N {\rm e}^{\gamma_E}$, and with 
\beqa
A (\alpha_s) &=& \nonumber\\
&\ & \hspace{-10mm}{\alpha_s\over \pi}\, C_F\, 
\left[ 1+  {\alpha_s\over 2\pi} \left( C_A\left({67\over 18}-
{\pi^2\over 6} \right)-{10\over 9}T_F\right)\right] 
\nonumber\\
B (\alpha_s) &=& {3\over 2}C_F{\alpha_s\over 2\pi}\, \, .
\label{ABdef}
\eeqa
Eq.\ (\ref{Edisdef}) is accurate to leading (LL) and next-to-leading 
logarithms (NLL)
in $N$ in the exponent: $\alpha_s^m\ln^{m+1}N$ and $\alpha_s^m\ln^mN$,
respectively.   The $N$ dependence of the ratio
$\tilde{C}_2^{\rm res}(N,Q,Q) / \tilde{C}_2^{\rm NLO}(N,Q,Q)$
is shown in Fig.\ \ref{cratio},
with $Q^2$ = 1, 5, 10, 100 GeV$^2$.
At $N=1$ the ratio is unity.  It is less than unity for
moderate $N$, but then begins to rise, with a slope that
increases strongly for small $Q$.   At low $Q^2$
and large $N$, higher orders can be quite important.
What does this  mean for PDFs?   We can certainly refit
PDFs with resummed coefficient functions, and 
we see that the high moments of such PDFs are likely
to be quite different from those from NLO fits.  

To get a sense of how such an NLL/NLO-$\overline{\rm MS}$ scheme
might differ from a classic NLO-$\overline{\rm MS}$
scheme, we resort to a model set of resummed
distributions, determined as follows.
We define valence PDFs in the resummed scheme
by demanding that their contributions to $F_2$ match those of the 
corresponding NLO
valence PDFs at a fixed $Q=Q_0$, which is ensured by
\beq
\tilde{\phi}^{\rm res}(N,Q_0^2)=\tilde{\phi}^{\rm NLO}(N,Q_0^2)\;
{ \tilde{C}_2^{\rm NLO}(N,Q_0,Q_0) \over 
\tilde{C}_2^{\rm res}(N,Q_0,Q_0)}\, .
\label{fresdef}
\eeq
Using the resummed parton densities from Eq.\ (\ref{fresdef}),
we can generate the ratios 
$F_2^{\rm res}(x,Q)/F_2^{\rm NLO}(x,Q)$.

\begin{figure}[t]
\epsfig{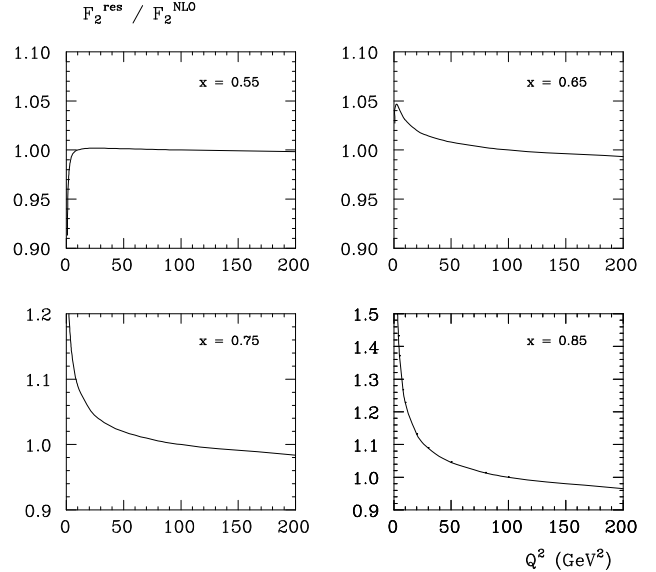}
\vspace{-0.8cm}
\caption{Ratio of the valence parts of the resummed and NLO proton 
structure function $F_2 (x,Q^2)$, as a function of $Q^2$ for various
values of Bjorken-$x$. For $F_2^{\rm res}$, the `resummed' parton densities 
have been determined through Eq.~(\ref{fresdef}).}
\label{fratio}
\end{figure}

The result of this test, picking $Q_0^2=100$ GeV$^2$
is shown in Fig.\ \ref{fratio}, for the valence $F_2(x,Q)$ of
the proton, with $x$ = 0.55, 0.65, 0.75 and 0.85.
The NLO distributions were those
of \cite{grv98}, and the inversion of moments
was performed as in \cite{minimal}.  The effect of
resummation is moderate for most $Q$.
At small values of $Q$, and large $x$, the resummed
structure function shows a rather sharp upturn. 
One also finds a gentle decrease toward very large $Q$ \cite{CM}.
We could interpret this difference as the
{\it uncertainty} in the purely NLO valence PDFs 
implied by resummation.

 From this simplified example, we can already see that the
use of resummed coefficient functions is not likely to make
drastic differences in global fits to PDFs based on DIS data, at least
so long as the region of small $Q^2$, of 10 GeV$^2$ or below, is avoided
at very large $x$.  At the same time, it is clear that a resummed
fit will make some difference at larger $x$, where PDFs are
not so well known. We stress that a full global fit
will be necessary for complete confidence.

\section{RESUMMED HADRONIC SCATTERING}

Processes other than DIS play an important role
in global fits, and in any case are of paramount phenomenological interest.
Potential sources of large corrections can be identified
quite readily in Eq.\ (\ref{momentact}).  At higher orders, 
factors such as $\alpha_s\ln^2N$, can
be as large as unity over the physically relevant range of 
$z$ in some processes.
In this case, they, and their scale dependence can be competitive with
NLO contributions.
Since they make up well-defined parts of the correction at each higher
order, however, it is possible to resum them.  
To better determine PDFs in regions of phase space where
such corrections are important, we may
incorporate resummation in the hard-scattering functions that determine
PDFs.

The Drell-Yan cross section is the benchmark for
the resummation of logs of $1-z$, or equivalently, logarithms of the
moment variable $N$ \cite{oldDY},
\beqa
\hat{\sigma}_{q\bar q}^{\rm DY}(N,Q,\mu) &=& \sigma_{\rm Born}(Q)\;
C_{\delta}^{\rm DY} \, {\rm e}^{E_{\rm DY}(N,Q,\mu)}
\nonumber \\
&\ & \hspace{10mm} + {\cal O}(1/N)\, .
\label{dyexp}
\eeqa
The exponent is given in the $\overline{\rm MS}$ scheme by
\beqa
E_{\rm DY}(N,Q,\mu) 
&=&  2\int^{\mu^2}_{Q^2/\bar{N}^2}{d\mu'{}^2\over\mu'{}^2}
A(\alpha_s(\mu'{}^2))\ln\bar{N}
\nonumber\\ 
&\ & \hspace{-15mm}
+ 2\int^{Q^2}_{Q^2/\bar{N}^2}{d\mu'{}^2\over\mu'{}^2}
A(\alpha_s(\mu'{}^2))\ln\left({\mu'\over Q}\right),
\label{Edydef}
\eeqa
with $A$ as in Eq.\ (\ref{ABdef}), and where we have
exhibited the dependence on the factorization scale,
setting the renormalization scale to $Q$.  Just as in Eq.\ (\ref{Edisdef})
for DIS,
Eq.\ (\ref{Edydef}) resums all leading and next-to-leading 
logarithms of $N$.

It has been noted
in several phenomenological applications that threshold resummation,
and even fixed-order expansions based upon it,
significantly reduce sensitivity to the factorization scale
\cite{scalered}.
To see why, we
rewrite the moments of the Drell-Yan cross section 
in resummed form as
\beqa
\sigma^{\rm DY}_{AB}(N,Q)
&& \nonumber\\
&& \hspace{-20mm}
= \sum_q\;
  \phi_{q/A}(N,\mu)\;
\hat{\sigma}_{q\bar q}^{\rm DY}(N,Q,\mu)\;
\phi_{\bar q/B}(N,\mu)
\nonumber\\
&& \nonumber\\
&& \hspace{-20mm} =
\sum_q\;
\phi_{q/A}(N,\mu)\; e^{E_{\rm DY}(N,Q,\mu)/2}
\sigma_{\rm Born}(Q) \, \, C_{\delta}^{\rm DY}
\nonumber\\
&& \hspace{-15mm} \times\;
\phi_{\bar q/B}(N,\mu)\; e^{E_{\rm DY}(N,Q,\mu)/2} + {\cal 
O}({1/ N})\, .
\eeqa
The exponentials compensate for the $\ln N$ part of the
evolution of the parton distributions, and the
  $\mu$-dependence of the resummed expression
is suppressed by a power of the moment variable,
\beq
\mu{d\over d\mu}\left[\, \phi_{q/A}(N,\mu)\; e^{E_{\rm 
DY}(N,Q,\mu)/2}\; \right]
=   {\cal O}({1/ N})\, .
\eeq
This surprising relation holds because the function
$A(\alpha_s)$ in Eq.\ (\ref{ABdef}) equals the
residue of the $1/(1-x)$ term in the splitting function $P_{qq}$.
Thus, the remaining $N$-dependence in a resummed cross section
still begins at order $\alpha_s^2$, but the part associated
with the $1/(1-x)$ term in the
splitting functions has been canceled to all orders.
Of course, the importance of the remaining sensitivity to $\mu$
depends on the kinematics and the process.   In addition,
although resummed cross sections can be made independent
of $\mu$ for all $\ln N$,
they are still uncertain at next-to-next-to leading logarithm
in $N$, simply because we do not know the function $A$ at three loops.
Notice that none of these results depends on using PDFs
from a resummed scheme, because
$\overline{\rm MS}$ PDFs, whether resummed or NLO, evolve the same way.
The remaining, uncanceled dependence on the scales leaves room for an educated
use of scale-setting arguments \cite{scales}.  The connection
between resummation and the elimination of scale
dependence has also been emphasized in \cite{maxwell}.

Scale dependence aside, 
can we in good conscience combine resummed hard scattering
functions in Eq.\ (\ref{basicfact}) with PDFs from an NLO scheme?
This wouldn't make much sense if resummation
significantly changed the coefficient functions
with which the PDFs were originally fit.
As Fig.\ \ref{fratio}
shows, however, this is unlikely to be the case for DIS at moderate $x$.
Thus, it makes sense to apply threshold resummation with NLO PDFs
to processes and regions of phase space where there is reason
to believe that logs are more important at higher orders than
for the input data to the NLO fits.

At the same time, a set of
fits that includes threshold resummation in their hard-scattering
functions can be made \cite{Cea1PI}, and their comparison to
strict NLO fits would be quite interesting.  Indeed, such a comparison
would be a new measure of the influence of higher orders.
A particularly interesting example might be to compare resummed and NLO
fits using high-$p_T$ jet data \cite{cteq5}.

\section{POWER-SUPPRESSED CORRECTIONS}

In addition to higher orders in $\alpha_s(\mu^2)$,
Eq.\ (\ref{basicfact}) 
has corrections that fall
off as powers of the hard-scattering scale $Q$.
In contrast to higher orders, these
corrections require a generalization of the {\it form}
of the factorized cross section.  
Often power corrections are parameterized
as $h(x)/[(1-x)Q^2]$ in inclusive DIS, where
they begin at twist four.
In DIS, this higher twist term influences PDFs when
included in joint fits
with the NLO and NNLO models, and vice-versa 
\cite{bodekyang,kateevetal,vanneervenvogt}.
As in the case with higher orders, such ``power-improved" fits
should be treated as new schemes.

\section{CONCLUSIONS}

The success of NLO fits to DIS and
the studies of resummation above
suggest that over most of the range of $x$, 
theoretical uncertainties of the NLO model
are not severe.   At the
same time, to fit large $x$
with more confidence than is now possible may
require including the 
resummed coefficient functions.

Resummation is especially desirable for
global fits that employ a variety of processes, such
as DIS and high-$p_T$ jet production,
which differ in available phase space near partonic
threshold.
In a strictly NLO approach, uncalculated large
corrections are automatically incorporated in the
PDFs themselves.
As a result, the NLO model cannot
be expected to fit simultaneously the
large-$x$ regions of processes with differing logs
of $1-x$ in their hard-scattering functions, unless
these higher-order corrections are taken into account.

The results illustrated
in the figures suggest that these
considerations may be important in
DIS with $Q^2$ below a few GeV$^2$ and at large $x$, where
they may have substantial effects on estimates
of higher twist in DIS.
In hadronic scattering, large-$N$ ($x\rightarrow 1$) resummation, which
automatically reduces scale dependence, may play
an even more important role than in DIS.

\subsection*{Acknowledgments}
We thank Andreas Vogt and Stephane Keller for useful discussions.

\end{document}